# K8s Pro Sentinel: Extend Secret Security in Kubernetes Cluster

Kavindu Gunathilake
Department of Computer Science
Informatics Institute of Technology
Sri Lanka
kavindu.2019822@iit.ac.lk

Indrajith Ekanayake
Department of Computer Science
Informatics Institute of Technology
Sri Lanka
indrajith.e@iit.ac.lk

*Abstract*— **Microservice architecture is widely adopted among distributed systems. It follows the modular approach that decomposes large software applications into independent services. Kubernetes has become the standard tool for managing these microservices. It stores sensitive information like database passwords, API keys, and access tokens as Secret Objects. There are security mechanisms employed to safeguard these confidential data, such as encryption, Role Based Access Control (RBAC), and the least privilege principle. However, manually configuring these measures is time-consuming, requires specialized knowledge, and is prone to human error, thereby increasing the risks of misconfiguration. This research introduces K8s Pro Sentinel, an operator that automates the configuration of encryption and access control for Secret Objects by extending the Kubernetes API server. This automation reduces human error and enhances security within clusters. The performance and reliability of the Sentinel operator were evaluated using Red Hat Operator Scorecard and chaos engineering practices.**

## I. Introduction

The term "microservices" was coined in 2011 at a software architecture workshop, which is built on the concept of Service-Oriented Architecture (SOA). This architecture is designed with three key concepts: flexibility, modularity, and maintainability[1], [2]. Initially, major industry players, including Netflix [1], Amazon[3], Meta[4], and Google [5], adopted this architecture. However, as their applications scaled, they encountered an exponential increase in complexity, which led to the adoption of container orchestrators such as Kubernetes, Docker Swarm, Mesos, and Red Hat OpenShift [6]. Kubernetes was initially developed and open-sourced by Google in 2014 [7]. Due to its open-source nature and capability to manage complex distributed systems, it has become a dominant tool for container orchestration.

Given the increasing complexity inherent in modular architectures, ensuring security has become a significant challenge. Kubernetes offers advanced security mechanisms designed to protect applications from attack vectors [8]. Nevertheless, Kubernetes is not secure by default; it is incumbent upon developers to follow best practices and enable advanced security features as necessary. For instance, Kubernetes stores confidential data within Secret Objects. By default, however, the Kubernetes API Server does not encrypt these Secrets; instead, it stores them as base64-encoded values, which do not provide true encryption and can be easily decoded [9]. Relying on this human-machine interaction increases the potential for human error, which highlights the necessity for automated security measures [10], [11], [12].

Addressing the gap of human error and misconfigurations in Kubernetes Secrets management, this study proposes an open-source framework (K8s Pro Sentinel) that is specialized for handling Secret deployments. The objective of this research is to extend the default Kubernetes functionality by adding automated Secrets management to the cluster including:

- To introduce a novel cloud-agnostic solution to automate security checks for Secrets management. (i.e., Role Based Access Control (RBAC) policy analysis, Secret encryption, etc.)
- To implement auditing through the Secret deployable user interface.
- To introduce an intuitive interface that does not require specialized knowledge of Kubernetes.

This publication is structured into several sections. Section II reviews the relevant literature and background that guided the development of the security optimization model. Section III details the methods employed to construct the proposed model, along with an overview of its key elements. Section IV analyzes the results derived from the model's application. Section V discusses the conclusions of this study and suggests directions for future research. Finally, section VI offers instructions to access the complete source code of the proposed solution.

## II. Related Work

Kubernetes cluster security encompasses four critical aspects, as outlined in the official documentation. These include control plane protection, Secrets protection, workload protection, and auditing [13]. Together, these elements contribute to determining the overall security posture of the cluster. While there are many attempts to safeguard each [14], [15], [16], this study only focuses on Secrets protection. Bueno and Block identify four distinct methods for securely integrating Secret within a cluster [17]. They are Role Based Access Control (RBAC), use external Key Management Service (KMS), encrypt the Secrets at rest (locally), and rotate the Secrets.

### *A. Manage Secrets with RBAC*

Kubernetes supports the deployment of Secrets with the etcd key-value store [18], employing base64 encoding as its default storage method. There are attempts to tailor this default behavior by implementing RBAC configurations [19]. However, this is not a cloud-agnostic solution, is particularly designed for the Amazon Elastic Kubernetes Service (EKS) leading to a void in other managed service clusters or local clusters.

Managing Secrets through RBAC allows administrators to define specific roles that are authorized to access the namespace where the Secrets are deployed. Such measures enhance security by mitigating unauthorized access to sensitive credentials within the cluster.

## B. Manage Secrets with external KMS

Kubernetes provides native integration with external KMSs for stringent security requirements such as banking systems. KMSs ensure that only encrypted values are stored within the etcd key-value store, thus bolstering the security of the data lifecycle management processes [20]. A study in 2019, identified 54 KMSs, out of which HashiCorp Vault, Ansible Vault, Google Cloud KMS, AWS Key Management Service (KMS), and Azure Key Vault are some of the most widely adopted solutions [21]. It concludes by mentioning HashiCorp Vault as the most suited KMS out of the 54 identified. However, most of these KMSs are expensive, which leads to a considerable void for small businesses to look for alternative options for secure Secrets management.

## C. Encrypt Secrets at rest (locally)

This strategy involves managing the encryption process which is not enabled by default. Implementing this requires configuring an “EncryptionConfiguration” object that specifies which Secrets to encrypt and the encryption provider [17]. This approach ensures that Secrets are stored securely, thereby enhancing the security of the Secrets when they are not actively being accessed or transmitted. However, this approach also brings logistical complexity when projects scale over time.

## D. Rotate the Secrets

Rotating Secrets involves replacing existing encryption keys with new ones to minimize the risks associated with key compromise[17]. Rotating Secrets helps to reduce the attack surface by limiting the time window an attacker has to exploit a stolen Secret. The necessity of Secret rotation is underscored by the static nature of most Secrets, which, if not rotated, can become easy targets for cyber-attacks. Various tools facilitate Secret rotation across different platforms. Ansible Vault uses a rekey functionality to allow admins to change passwords secure encrypted files, and maintain the integrity of stored data [22]. Sealed Secrets for Kubernetes automates key renewals and allows manual key rotation, keeping Kubernetes Secrets secure in dynamic environments [17]. SOPS (Secrets OPerationS), used with Helm Secrets, supports comprehensive key rotations, including both data and master GPG keys, ensuring high levels of security in critical environments [23].

In Kubernetes, although the architecture supports Secret rotation, the process can be complex due to the interconnected nature of services and their reliance on consistent Secret access. Hence, there is more room for misconfigurations which could mitigated by automation and minimizing human involvement.

According to Akon Rahman, >=15.7% of Kubernetes manifests exhibit one or more misconfigurations [24]. To come to this conclusion, the study has used 2,039 Kubernetes manifests from open-source repositories. It’s not because Kubernetes is faulty, Kubernetes is designed to loosely manage security to accommodate "vanilla" deployments, aiming to serve a broad range of users. The same problem is further underscored in many studies that highlight the importance of paying greater attention to security [10], [11], [12]. This misconfiguration and human error can make the clusters vulnerable [24]. Previous works identify automation as the solution to this problem. Furthermore, some studies have specifically suggested the use of Kubernetes operators for archiving automation[25].

# III. Methodology

To address the gap identified in Kubernetes Secrets misconfiguration, the proposed model, K8s Pro Sentinel, is created by extending the Kubernetes API server as a Custom Resource (Fig 1). It provides a mechanism to add customized logic to a running Kubernetes cluster [26]. Once installed, users can create and interact with its objects similarly to how they manage standard resources such as Pods and Deployments. There are two types of Custom Resources, which are Custom Resource Definitions (CRD) and API aggregation. K8s Pro Sentinel uses CRD with a controller, which provides a reconcile function responsible for synchronizing Secret resources until the desired state is reached.

Fig 2 presents the three-tier architecture used to design the K8s-Pro Sentinel. This architecture pattern, widely adopted in software development, segregates an application into three interrelated components, each tasked with distinct functionalities [27].

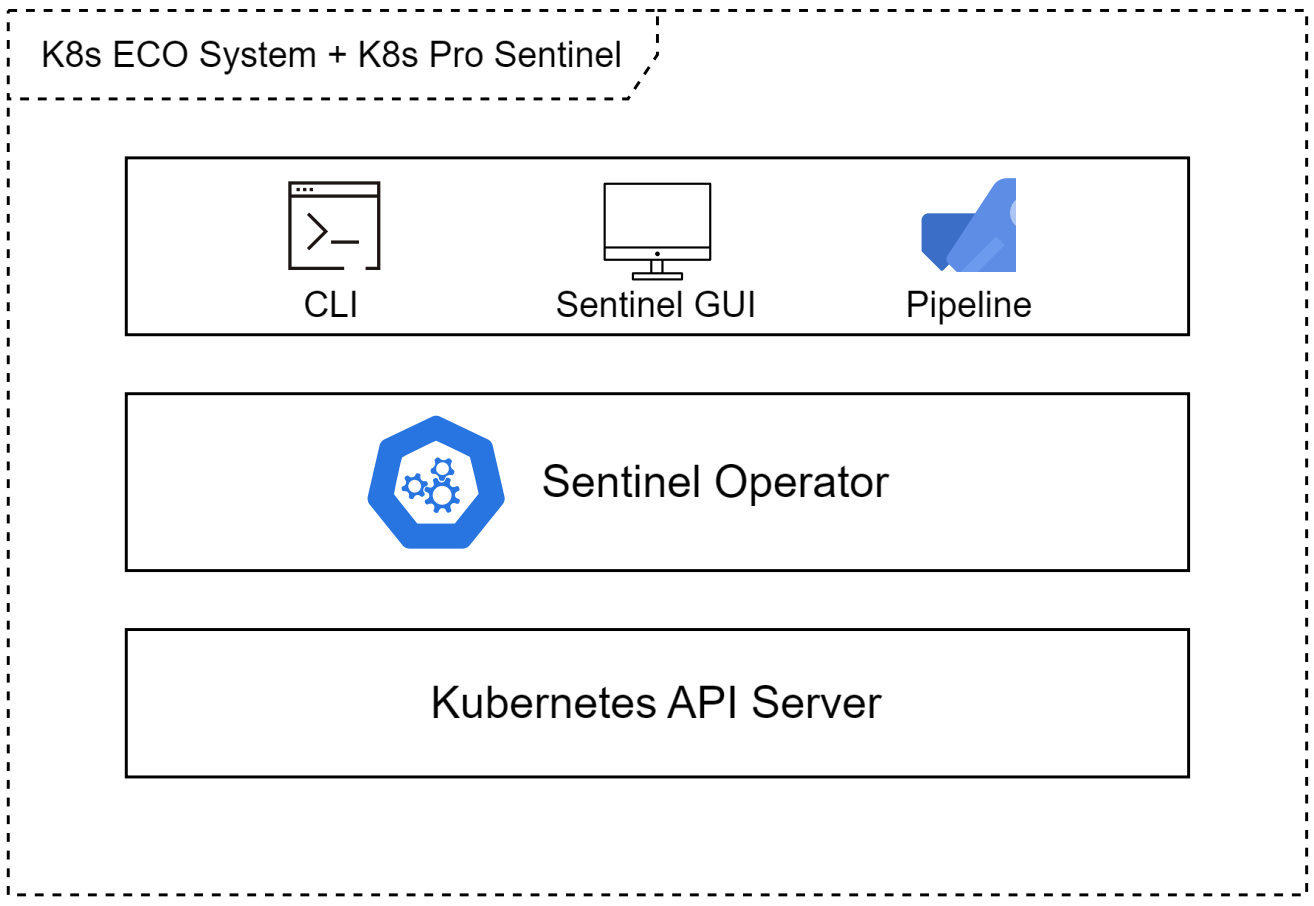

Fig. 1. Sentinel operator extending Kubernetes API server

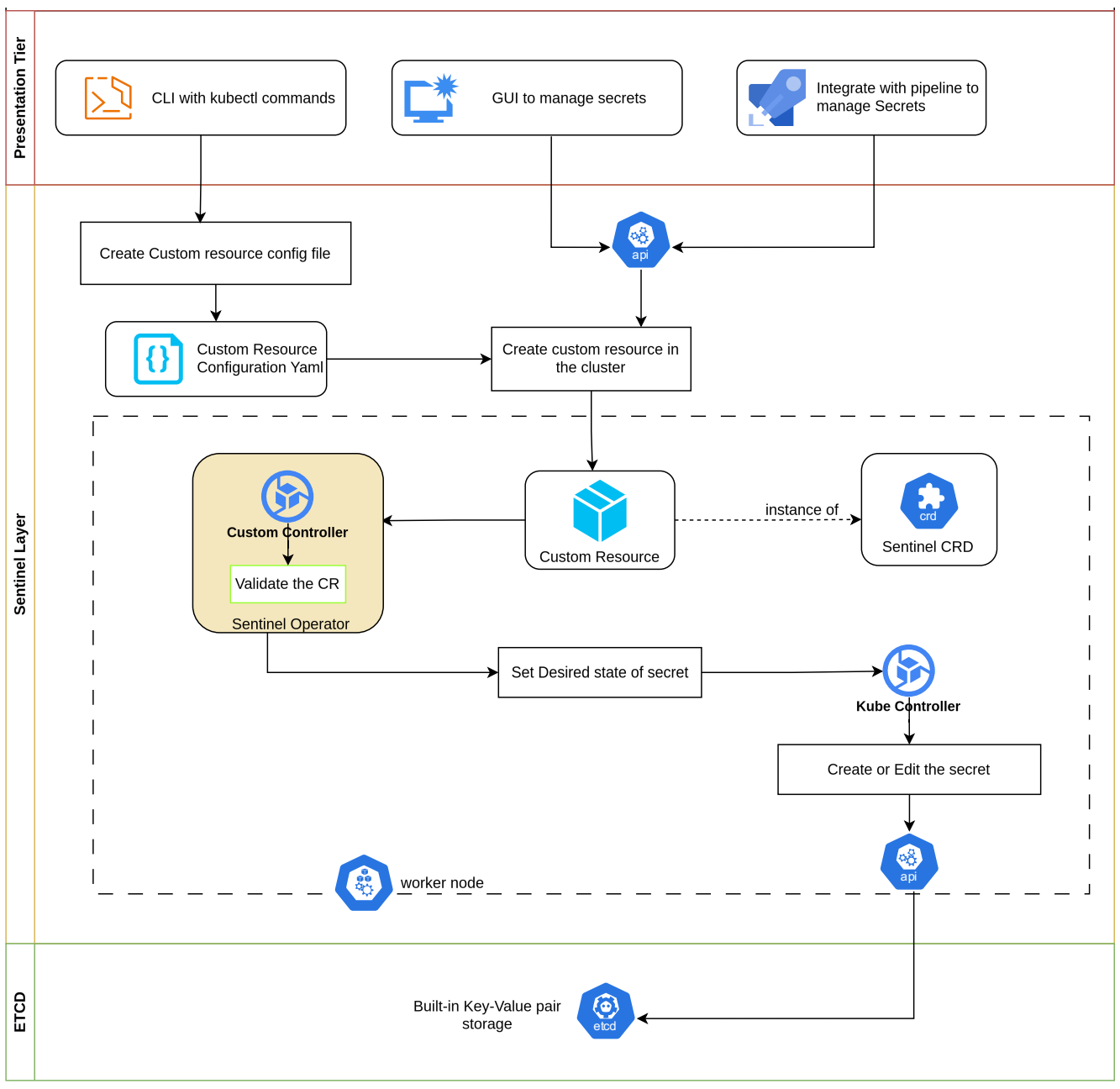

Fig. 2. The system architecture of K8s Pro Sentinel

The presentation tier includes both a Command Line Interface (CLI) and a Graphical User Interface (GUI), which enable end users to create and modify resources. For instance, by following the templates, the user can manually use the "kubectl" command to create the Custom Resource. Alternatively, the user can leverage the GUI to fill out forms that apply the Custom Resource to the cluster. This interface also provides error notifications in case of any deployment errors.

```yaml
---
apiVersion: apiextensions.k8s.io/v1
kind: CustomResourceDefinition
metadata:
  annotations:
    controller-gen.kubebuilder.io/version: v0.12.0
  name: sentinels.secops.kavinduxo.com
spec:
  group: secops.kavinduxo.com
  names:
    kind: Sentinel
    listKind: SentinelList
    plural: sentinels
    singular: sentinel
  scope: Namespaced
```

Fig. 3. CustomResourceDefinition (CRD) configuration - this is a snapshot of the first 15 lines, for the complete YAML file refer to the K8s Pro Sentinel open-source project.

The middle tier is composed of proxy API that facilitates communication with the GUI and various pipelines. This API is responsible for creating Custom Resources and delivering the resultant messages. The Sentinel CRD serves as the blueprint for these Custom Resources (Fig 3). Custom Resources are generated within the cluster based on the CRD and user inputs. The central component in the middle tier is the Sentinel Operator, which incorporates a custom controller. This controller validates the Custom Resource by checking the types and preconditions for creating or modifying Secrets in the cluster. The Controller Manager oversees the custom controller's reconciliation processes, while the standard Kubernetes controller ensures that the state of the Secret is reconciled to match the desired configuration and implements any required changes.

The database tier houses an etcd key-value store, which maintains data in key-value pairs. When modifications are made to a Secret, this storage system records the Secret's information, which can be either encoded or encrypted using the EncryptionProviderConfig type Secret.

The proposed solution minimizes human error and misconfiguration through automation and introducing a GUI where users can create and manage Secrets with guided steps. K8s Pro Sentinel consists of these functionalities:

- Create BaseSecret ("LocalEncryptionProvider" Secret type, or RBAC type) through CLI or GUI. RBAC Secrets validated with the service account, role binding, and role.
- Modify BaseSecret or RBAC type Secret either through CLI or GUI.
- Automated auditing of created/ existing Secrets through the CLI or GUI.

When it comes to technologies used to develop the solution, The controller/ operator logic is implemented using operator SDK and Go language, which is a common language of choice when it comes to writing custom controller logic [28]. The user interface is created using ReactJS. The operator was developed in an Ubuntu operating system with git as the version-controlling mechanism, VS Code as the primary IDE, and Docker Desktop was used for running the local 2-Node Kubernetes cluster.

## IV. Results and Discussion

For evaluating the proposed architecture, the authors used a self-hosted Kubernetes cluster using Docker Desktop with one worker node and one master node. "Makefiles" were used for automating the building and deploying process of the operator. The latency was captured while installing the operator, deploying Custom Resource, and validation of Secret Objects both using CLI (Fig 4), and GUI (Table I).

```
[16:56:36] $kubectl apply -f config/samples/secops_v1alpha1_sentinel_matrix.yaml
ecops.kavinduxo.com/base-rbac-sentinel unchanged
[16:56:37] $
```

Fig. 4. Mesure latency while creating RBAC type Secret object.

TABLE I. Latency (Measured in Seconds)

| ***Function*** | ***From GUI*** | ***From CLI*** |
|---|---|---|
| Installing operator | N/A | 4s |
| Deploying Custom Resource | 2s | 1s |
| Secret validation | 3s | 2s |

The evacuation of creating and modifying BaseSecret, LocalEncryptionProvider Secret, and RBAC type Secret was conducted using Chaos experiments [29]. Following chaos engineering principles, authors intentionally introduced failures into the system (e.g., introducing Secrets that are not secured to the cluster, or Create Secrets without following best practices). This facilitated an understanding of how automated auditing reacts to changes within the cluster and addresses faults. Finally, the quality and adherence to the best practices of the proposed Custom Resource were validated using the Operator SDK Scorecard [30]. While it's possible to execute custom test definitions, authors have only used built-in basic and Operator Lifecycle Manager (OLM) tests. There was a total of six tests including:

*1) Spec block exist:* Make sure every Custom Resource contains the correct spec in the body.

```
Image:      quay.io/operator-framework/scorecard-test:v1.33.0
Entrypoint: [scorecard-test basic-check-spec]
Labels:
        "suite":"basic"
        "test":"basic-check-spec-test"
Results:
        Name: basic-check-spec
        State: pass
```

Fig. 5. Spec block exist test result

*2) Bundle validation:* Validates bundle manifests, identifies content errors, and suggests corrections.

```
Image:      quay.io/operator-framework/scorecard-test:v1.33.0
Entrypoint: [scorecard-test olm-bundle-validation]
Labels:
        "suite":"olm"
        "test":"olm-bundle-validation-test"
Results:
        Name: olm-bundle-validation
        State: pass

        Suggestions:
```

Fig. 6. Bundle validation test result

*3) Exposed API validation:* Confirms that the CRDs include a validation section for all spec and status fields, ensuring data integrity.

```
Image:      quay.io/operator-framework/scorecard-test:v1.33.0
Entrypoint: [scorecard-test olm-crds-have-validation]
Labels:
        "suite":"olm"
        "test":"olm-crds-have-validation-test"
Results:
        Name: olm-crds-have-validation
        State: pass

        Log:
                Loaded 1 Custom Resources from alm-examples
```

Fig. 7. Exposed API validation test result

*4) Owned CRDs have resources listed:* Checks that CRDs list all utilized resources in their resources subsection, ensuring comprehensive tracing and documentation.

```
Image:      quay.io/operator-framework/scorecard-test:v1.33.0
Entrypoint: [scorecard-test olm-crds-have-resources]
Labels:
        "test":"olm-crds-have-resources-test"
        "suite":"olm"
Results:
        Name: olm-crds-have-resources
        State: fail

        Errors:
                Owned CRDs do not have resources specified
        Log:
                Loaded ClusterServiceVersion: sentinel-operator.v0.0.1
```

Fig. 8. Owned CRDs have resources listed test result

*5) Spec fields with descriptors:* Ensures each field in the spec sections of Custom Resources includes a descriptor in the Cluster Service Version (CSV).

```
$operator-sdk scorecard ./bundle -o text --selector=suite=olm
--------------------------------------------------------------------------------
Image:      quay.io/operator-framework/scorecard-test:v1.33.0
Entrypoint: [scorecard-test olm-spec-descriptors]
Labels:
        "suite":"olm"
        "test":"olm-spec-descriptors-test"
Results:
        Name: olm-spec-descriptors
        State: pass

        Log:
                Loaded ClusterServiceVersion: sentinel-operator.v0.0.1
                Loaded 1 Custom Resources from alm-examples
```

Fig. 9. Spec fields with descriptors test result

*6) Status fields with descriptors:* Confirms that every field in the status sections of Custom Resources is paired with a descriptor in the Cluster Service Version (CSV) for accurate status reporting.

```
Image:      quay.io/operator-framework/scorecard-test:v1.33.0
Entrypoint: [scorecard-test olm-status-descriptors]
Labels:
        "suite":"olm"
        "test":"olm-status-descriptors-test"
Results:
        Name: olm-status-descriptors
        State: pass

        Suggestions:
```

Fig. 10. Status fields with descriptors test result

TABLE II. SUMMARY OF OPERATOR SDK SCORECARD TESTING

| ***Test Case*** | ***Test Result*** |
|---|---|
| Spec block exist | Pass |
| Bundle validation | Pass |
| Exposed API validation | Pass |
| Owned CRDs have resources listed | Pass |
| Spec fields with descriptors | Fail |
| Status fields with descriptors | Pass |
| Scorecard standard tests pass rate = 5/6 * 100% = 83.3% | |

## V. CONCLUSION

This study addresses vulnerabilities in Kubernetes by mitigating human error and misconfiguration of Secret Objects through the novel open-source tool, K8s Pro Sentinel. This tool extends the Kubernetes API server, offering a cloud-agnostic mechanism for the automated management and auditing of Kubernetes Secrets. The empirical evaluation demonstrates the performance of K8s Pro Sentinel in enhancing security practices within a self-hosted Kubernetes environment.

Nevertheless, the scope of this evaluation was limited to a self-hosted setup, underscoring the necessity for further testing across various managed Kubernetes services to ensure broader applicability and reliability in diverse operational contexts. Furthermore, the existing academic literature lacks comprehensive studies in this domain, indicating a substantial opportunity for future research. Future studies could explore the extension of the Kubernetes API server functionality to strengthen the overall security posture of Kubernetes clusters.

## VI. SOURCE CODE

The source code for the proposed solution can be obtained from https://github.com/kavinduxo/k8s-pro-sentinel